\documentclass[12pt]{article}
\usepackage{amsfonts}
\usepackage{amssymb}
\usepackage{mathrsfs}
\usepackage{amsthm}
\usepackage{graphicx}
\usepackage{amsmath}
 \usepackage[T1]{fontenc} 

\newtheorem{thm}{Theorem}

\newtheorem{lemma}{Lemma}
\theoremstyle{definition}

\newtheorem{fact}{Fact}[section]

\newcommand{\bei}{\begin{itemize}}
\newcommand{\eei}{\end{itemize}}

\begin{document}

\vspace{5cm}

\begin{center}

{\Large DISSECTING THE QUTRIT}

\vspace{1cm}

{\large Gniewomir Sarbicki}\footnote{gniewko@fizyka.umk.pl}

\vspace{1cm} 

{\sl Fysikum, Stockholms Universitet,} 

{\sl S-106 91 Stockholm, Sweden}

{\sl Institute of Physics, Nicolaus Copernicus University,} 

{\sl Grudzi\k{a}dzka 5/7, 87-100 Toru\'{n}, Poland}

\vspace{1cm}

{\large Ingemar Bengtsson}\footnote{ibeng@fysik.su.se}

\vspace{1cm} 

{\sl Fysikum, Stockholms Universitet,} 

{\sl S-106 91 Stockholm, Sweden}

\vspace{1cm}

{\bf Abstract}

\vspace{1cm}

\end{center}

\noindent To visualize a higher dimensional object it is convenient to consider its 
two-dimensional cross-sections. The set of quantum states for a three level system 
has eight dimensions. We supplement a recent paper by Goyal et al by 
considering the set of all possible two-dimensional cross-sections of the qutrit. 
Each such cross-section is bounded by a plane cubic curve.

\newpage

\section{Introduction}

With the rise of quantum information theory, a minor issue but intriguing 
issue has attracted some attention: What does the set of quantum states 
actually look like? 

For a qubit the set of quantum states is a ball, but 
the case of a qutrit is already complicated \cite{Bl76}. It is an eight dimensional 
convex body. A standard way to visualize higher dimensional sets is to consider 
two-dimensional cross-sections, and preferably an intelligently organized 
Grand Tour of such 2-sections \cite{Asimov}. Of course this idea has been pursued for the 
qutrit, typically by considering the finite set of cross-sections generated 
by the Gell-Mann matrices \cite{Jakob} - \cite{BWZ}. An elegant and comprehensive 
treatment is due to Goyal et al \cite{Goyal}. However, by concentrating on 
cross-sections generated by the Gell-Mann matrices one obtains only 28 2-sections 
altogether, or 56 three-dimensional 3-sections. This is quite far from 
a Grand Tour. On the other hand we are looking at a fairly symmetric 
eight-dimensional body, 
left invariant by an $SU(3)$ subgroup of the rotation group $SO(8)$, and this 
cuts down the size of the problem. Our purpose here is to supplement 
the work by Goyal et al. by making this idea precise. In section \ref{GM} this 
will enable us to 
see how their selection samples the full set of 2-sections, because we 
will be able to visualize the set of all unitarily inequivalent two-dimensional cross-sections 
of the state space of the qutrit. In section \ref{gen-sec} we will consider 
the shape of a general 2-section, and the plane cubic curves that bound them. 

Because of the duality between sections and projections of a self-dual set, a 
description of two-dimensional cross-sections can be translated to results on 
the shapes of the numerical ranges of complex matrices \cite{Dunkl1}.
The numerical range is an interesting tool which has found some applications 
in quantum information theory \cite{Dunkl2}. This gives some special interest 
to the study of 2-sections, as opposed to 3-sections. 
We discuss this topic in section \ref{duality}.

\section{Preliminaries}

A 2-section of an object in an eight dimensional vector space is 
its intersection with a 2-plane through the origin---which will be placed at the 
maximally mixed qutrit state in our case. The modeling linear subspace for such a plane is a 2-dimensional subspace $V$ of the real vector space of traceless $3 \times 3$ hermitian matrices equipped with the Hilbert-Schmidt 
scalar product 
\begin{equation} M_1\cdot M_2 = \frac{1}{2}{\rm Tr}M_1M_2 \ . \end{equation} 
The set of quantum states is the convex set of traceless hermitian 
matrices $M$ such that the density matrix $\rho = {\bf 1}/3 + M$ is positive. 
It is inscribed in a minimal sphere of radius $R_{\rm out}$, and contains a 
sphere of maximal radius $R_{\rm in}$, where

\begin{equation} R_{\rm out} = \frac{1}{\sqrt{3}}, \hspace{8mm} R_{\rm in} = 
\frac{1}{2}R_{\rm out} \ . \end{equation}

We need to know the dimension of 
the set of all 2-sections, or equivalently of the set of all 2-dimensional 
subspaces of a real 8-dimensional space. This set is also known as the 
Grassmannian $Gr(2,6)$. Its dimension is easily found:

\begin{fact}
 The set of orthonormal bases in $\mathbb{R}^n$ is isomorphic to the orthogonal group $O(n)$.
\end{fact}

\begin{fact}
 The dimension of the group $O(n)$ is $\frac {n(n-1)}2$.
\end{fact}

\begin{fact}
 The set of $2$-dimensional subspaces of an $8$-dimensional real vector spaces has $12$ dimensions.
\end{fact}

\proof: Any 2-dimensional subspace can be obtained by choosing an orthonormal basis such 
that its first two vectors span the subspace, and the remaining six its orthogonal complement. But we obtain the same subspace if we change the bases within the subspace and within its 
complement. Hence the dimension of the set of 2-planes equals the dimension of $O(8)$ 
minus the dimension of the subgroup $O(2)\times O(6)$. So the answer is $8\cdot 7/2 - 
2\cdot 1/2 - 6\cdot 5/2 = 12$. 
$\square$

A general density matrix is given by 

\begin{equation} \rho = \frac{1}{3}I_3 + M \ , \end{equation}

\noindent where $I_3/3$ is the maximally mixed state and $M$ is a traceless 
hermitian matrix chosen such that all eigenvalues of $\rho$ are non-negative. 
The group $SU(3)$ acts on our vector space through $M \to UM U^\dagger$. This is an 8-dimensional 
subgroup of the rotation group $SO(8)$. Since the action of $SU(3)$ leaves the set of density matrices invariant all sections with subspaces related by this action will be regarded as equivalent. Hence we have arrived at: 

\begin{fact}
 The dimension of the set of inequivalent 2-sections of the set of states of the qutrit has 
 only $12 - 8 = 4$ dimensions.
\end{fact}

\noindent Equivalent 2-sections have the same shape, but the converse does not hold \cite{Goyal}.

\section{Representatives of 2-sections}

Let $V$ be a 2-dimensional subspace of the real vector space of traceless hermitian 
matrices. We begin with a simple observation:

\begin{lemma} The subspace $V$ contains no elements of rank $1$ and at least one element of rank $2$. \end{lemma}

\proof Assume that the subspace $V$ is spanned by elements $A$ and $B$. A non-zero element of $V$ is either of rank $2$ or of rank $3$ because is traceless. Then assume that $A$ is of rank $3$ (otherwise the proof is complete). Consider the determinant of a linear combination:
\begin{displaymath}
 \det(\lambda A+B)=0
\end{displaymath}
Because $A$ and $B$ are hermitian, this is a real polynomial of the order $3$. It has at least one real root. If $\lambda$ is this root the combination is of rank two. $\square$

Since we are interested in picturing the set of 2-sections up to unitary equivalence, we 
use Lemma 1 to introduce a standard form for a basis in the 2-plane. We choose 
the first basis vector $A$ to have rank 2, and then we choose the basis of $\mathbb{C}^3$ so that $A$ is diagonal. This fixes the basis in $\mathbb{C}^3$ up to diagonal unitary transformations, and we use the remaining freedom to adjust the phases of the orthogonal 
basis vector $B$. In this way we arrive at the following basis for the 2-plane:
\begin{align} 
 & A = 
 \left[\begin{array}{ccc}
 1 & 0 & 0 \\ 0 & -1 & 0 \\ 0 & 0 & 0
 \end{array}\right]
 && B = \left[\begin{array}{ccc}
 k & a e^{i \varphi} & b e^{i \varphi} \\ a e^{-i \varphi} & k & c e^{i \varphi} \\ b e^{-i \varphi} & c e^{-i \varphi} & -2k
 \end{array}\right] && \mathrm{if} \ abc \ne 0 \label{A} \\
 & A = 
 \left[\begin{array}{ccc}
 1 & 0 & 0 \\ 0 & -1 & 0 \\ 0 & 0 & 0
 \end{array}\right]
 && B = \left[\begin{array}{ccc}
 k & a & b \\ a & k & c \\ b & c & -2k
 \end{array}\right] && \mathrm{if} \ abc = 0 \label{B}
\end{align}
where we may assume that $a,b,c,k \ge 0$. We also define $d = \sqrt{3}k$. The 
matrices $A$ and $B$ form an orthonormal basis if and only if  
\begin{equation} 3k^2 + a^2 + b^2 + c^2 = a^2 + b^2 + c^2 + d^2 = 1 \ . \end{equation}
The set of possible tuples $(a, b, c, d)$ form 1/16 of a three dimensional sphere, and so it is homeomorphic to a three dimensional simplex. The set of matrices $B$ contains one representative of each unitary equivalence class of 2-sections of the set of qutrit states.  We can think of it as a Cartesian product of a simplex and a circle, except that the circles shrink to points at three of the four faces of the simplex.

Unfortunately some ambiguities remain because there may be more than one matrix 
of rank 2 in the 2-plane, and one can then perform a unitary transformation so that another matrix takes our standard form $A$. In particular given a matrix of rank 
2 its negative also has rank 2. If we leave the origin in the direction of a rank 2 
matrix we will hit the boundary of the set of states at a density matrix of spectrum $(2/3,1/3,0)$. Such a density matrix lies on a sphere of radius $R_{\rm 2}$, where 

\begin{equation} R_{\rm in} < R_{\rm 2} = \frac{1}{\sqrt{6}} < R_{\rm out} \ . \end{equation} 

\noindent Goyal et al \cite{Goyal} refer to this as the self-dual sphere, because 
a 2-section containing a point on this sphere will also contain its antipodal point. But this contradicts another and more common use of the word self-dual, to be 
discussed in section \ref{duality}.
Anyway the point is that the boundary of the set of quantum states intersects the sphere 
of radius $R_2$ in antipodal points. Now a permutation of the first and the second vector of the basis of the Hilbert space will change the sign of $A$ and exchange the parameters $b,c$ in $B$. Thus we are led to identify pairs of points related to each other by reflection with respect to the surface $b=c$. We reduce then the set of representatives to the halfsimplex HS of matrices $B$ where $b\ge c$ (times the circle, which shrinks to a point at two of the faces of the halfsimplex). 

To see if there are further ambiguities we solve the equation $\det (\lambda A + B)=0$. By definition of $A$ one solution is for $\lambda=\infty$. The other solutions are the roots of
\begin{equation} \label{dod_pierw}
 2k \cdot \lambda^2 + \lambda (b^2-c^2) + 2 abc \cos \phi -2k^3 + k(2a^2-b^2-c^2) \ . 
\end{equation}
First consider the situation when all elements of the subspace have rank 2. It happens if and only if $k=0, \ b=c$ and $abc\cos\phi=0$. The resulting 2-section is always a circular disk 
of radius $R_{\rm 2}$. As noted by Goyal et al (in a special case) these 2-sections are 
unitarily inequivalent even though they have the same shape. To see this, observe that 

\begin{equation} \frac{1}{2}\mbox{Tr}(AB-BA)^2 = 4a^2 +b^2 + c^2 =  3(a^2-k^2) + 1 
\ . \end{equation}

\noindent Once we have restricted ourselves to $k = 0$ it follows that the 
value of $a$ cannot be changed by changing the basis in the 2-plane $V$, so no 
ambiguities arise in this case.

If the discriminant of (\ref{dod_pierw}) is greater than or equal zero we have 
a discrete ambiguity in the choice of our matrix $A$, and this does give rise to 
discrete ambiguities in our parametrization. In particular one can check that 

\begin{equation} (k,a,b,c) = \left( 0,0,\cos{\frac{\theta}{2}}, \sin{\frac{\theta}{2}}\right) 
\hspace{5mm} \mbox{and} \hspace{5mm} (k,a,b,c,\phi ) = \left( \frac{\cos{\theta}}{2}, 
\frac{\cos{\theta}}{2},\frac{\sin{\theta}}{\sqrt{2}},\frac{\sin{\theta}}{\sqrt{2}}, 0\right) 
\end{equation}

\noindent correspond to equivalent 2-sections. In general the discriminant of 
eq. (\ref{dod_pierw}) vanishes if either $k = 0$ or a complicated $\phi$-dependent 
condition holds. We decided to ignore this difficulty.

We get the following:

\begin{thm}
 Any 2-dimensional section of the space of traceless hermitian matrices in $\mathbb{C}^3$ can be represented after appropriate change of basis of $\mathbb{C}^3$ as a subspace spanned by matrices of the form given in eqs. (\ref{A}-\ref{B}),
where $a^2+b^2+c^2 + 3k^2=1$, $b\ge c$ and $\phi$ is arbitrary. Topologically this set
forms a simplex times a circle, with the circles shrinking to points at two of 
the faces of the simplex (when $abc = 0$). The parametrization determines the set of 
unitarily equivalent 2-sections uniquely except for discrete ambiguities that occur if the 
2-section contains exactly 4 or exactly 6 traceless matrices of rank 2. 
\end{thm}

\begin{figure}[h!]
\begin{tabular}{rr}
\begin{minipage}{7.5cm}
 \centering
 \includegraphics[width=7cm]{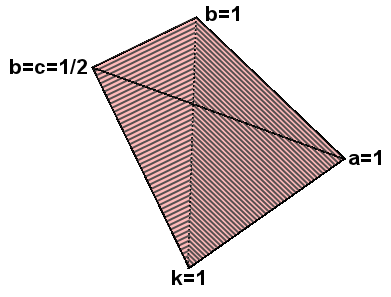}
 \caption{Halfsimplex. All points not contained in the dashed area (with its boundary) has an additional degree of freedom.
 $\phi$}
 \label{fig:tetr}
 \end{minipage}
\begin{minipage}{7cm}
\vspace{0.3cm}
\centering
 \includegraphics[width=5cm]{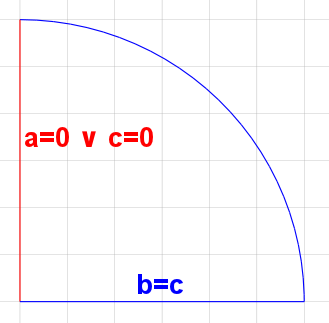}
 \caption{Halfsimplex deformed and represented in two dimensions. All points from the dashed area are marked as red.}
 \label{fig:tetr-red}
\end{minipage}
\end{tabular}
\end{figure}

The halfsimplex looks as in figure \ref{fig:tetr}. All points not contained in the dashed area (with its boundary) have an additional degree of freedom---the angle $\phi$. To visualize this 4-dimensional set, first deform the half-simplex homeomorphically (not diffeomorphically) such that all dashed area lie on one 2-dimensional surface, and the surface $b=c$ is orthogonal to it. Now reduce the dimension of such a deformed simplex as in figure \ref{fig:tetr-red}. 
Next we rotate in four dimensions using rotations that leave the surface $a=0 \lor c=0$ invariant. In the reduced picture this is related to rotation around the axis $y$. After rotation we get a half of a closed ball. The topology of resulting set is not interesting, however keep in mind, that to get one-to-one correspondence with the set of sections, some pairs of points should be glued together.


\section{Sections spanned by Gell-Mann matrices} \label{GM}
Goyal et al \cite{Goyal} consider 2-sections spanned by pairs of the eight Gell-Mann matrices, and illustrate them beautifully. 
The Gell-Mann matrices form an orthonormal basis of the vector space of traceless hermitian matrices, and are defined by 

\begin{equation} \sum_{i=1}^8x_i\lambda_i = \left[ \begin{array}{ccc} x_8/\sqrt{3} + x_3& 
x_1-ix_2 & x_4 - ix_5 \\ x_1+ix_2 & x_8/\sqrt{3} - x_3& x_6-ix_7 \\ x_4+ix_5 & 
x_6+ix_7 & -2x_8/\sqrt{3} \end{array} \right] \ . \end{equation}

\noindent Matrices $\lambda_1, \dots, \lambda_7$ are of rank two and are unitarily 
equivalent to our matrix $A = \lambda_3$. 

To check how a section spanned by a pair of Gell-Mann matrices is represented we bring one of them to the standard form $\lambda_3$ and call it $A$. The same operation brings then the other matrix to the form $B$, which is the representative of the section. Calculating it one gets (and we refer to Goyal et al. \cite{Goyal} for illustrations):
\begin{itemize}
 \item For pairs $12, 13, 23, 45, 67$ the representative is
 \begin{displaymath}
  \left[ \begin{array}{rrr} 0 & 1 & 0 \\ 1 & 0 & 0 \\ 0 & 0 & 0 \end{array} \right]
 \end{displaymath}
 and the shape of intersection is circular disk. As a side remark we observe that 
 the 4-section given by the quartet $1245$ (say) is a round ball.
 \item For pairs $14, 15, 16, 17, 24, 25, 26, 27, 46, 47, 56, 57$ the representative is
 \begin{displaymath}
  \frac 1{\sqrt{2}} \left[ \begin{array}{rrr} 0 & 0 & 1 \\ 0 & 0 & 1 \\ 1 & 1 & 0 \end{array} \right]
 \end{displaymath}
 and the shape of the intersection is a circular disk, unitarily inequivalent to 
 the above. 
 \item For pairs $34, 35, 36, 37$ the representative is
 \begin{displaymath}
  \frac 12 \left[ \begin{array}{rrr} 1 & 1 & 0 \\ 1 & 1 & 0 \\ 0 & 0 & 2 \end{array} \right]
  \hspace{5mm} \mbox{or} \hspace{5mm} \left[ \begin{array}{rrr} 0 & 0 & 1 \\ 0 & 0 & 0 \\ 1 & 0 & 0 \end{array} \right]
 \end{displaymath}
 and the shape of the intersection is a parabola. Note that the discrete ambiguity in our 
 parametrization turns up here.
 \item For pairs $18, 28, 38$ the representative is
 \begin{displaymath}
  \frac 1{\sqrt{3}} \left[ \begin{array}{rrr} 1 & 0 & 0 \\ 0 & 1 & 0 \\ 0 & 0 & -2 \end{array} \right]
 \end{displaymath}
 and the shape of the intersection is an equilateral triangle.
 \item For pairs $48, 58, 68, 78$ the representative is
 \begin{displaymath}
  \frac 1{2\sqrt{3}} \left[ \begin{array}{rrr} 1 & 3 & 0 \\ 3 & 1 & 0 \\ 0 & 0 & -2 \end{array} \right]
 \end{displaymath}
 and the shape of the intersection is an ellipse.
 
All these cases have $\cos\varphi=0$ and sit at the edges of the halfsimplex, as shown (together with their shapes)  in figure \ref{fig:tetr1}.
 \begin{figure}[h!]
 \centering
 \includegraphics[width=10cm]{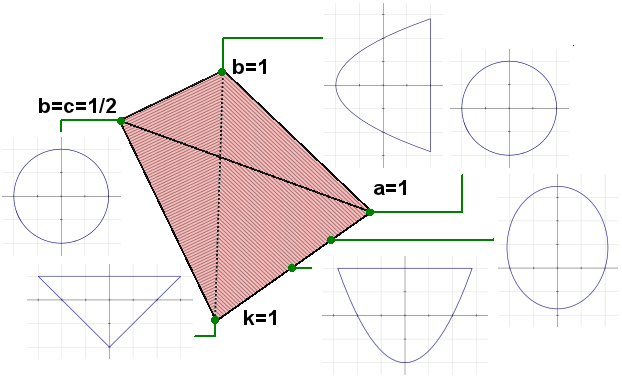}
 \caption{The shapes of special sections and their positions in the halfsimplex $\varphi=0$.}
 \label{fig:tetr1}
\end{figure}

\end{itemize}

\section{The shape of a general 2-section} \label{gen-sec}
So far we have sampled only some very special points in the set of all 
2-sections. As we have seen, up to unitary transformations it is enough 
to look at density matrices of the form $\rho = I_3/3 + x A + y B$. The boundary 
of a 2-section is described by

\begin{equation} \label{cub-cond}
   3\det{\rho} = \frac 19 - x^2 - y^2  + 3(2abc\cos \varphi-k(1-k^2-3a^2))\cdot y^3 + 6k\cdot x^2y + 3(b^2-c^2)\cdot x y^2 = 0 \ .  
\end{equation}

\noindent The problem of classifying all 2-sections is thereby reduced to the 
problem of classifying plane cubic curves of a somewhat special form. 
Here we will be concerned with two questions: When does the cubic curve 
factorize into three linear factors or into one linear and one quadratic factor? 
If it does not factorize, when is the boundary of the 2-section not smooth?
  
The cubic factorizes in the following cases:
\begin{enumerate}
 \item $b=c \ \land \ (b=0 \lor a \cos \varphi = 3k)$. There are two families of sections. One lies on the boundary $b=c=0$ of the simplex and the second forms for any $\phi \in (-\frac \pi 2, \frac \pi 2)$ a line starting from the point $b=c=1/\sqrt{2}$ and passing the surface $b=c$. For $\phi \in (\frac \pi 2, 3 \frac \pi 2)$ the second family is only one point $b=c=1/\sqrt{2}$. The 2-section is described by 
 \begin{equation}
  \left(\frac 13 - 2ky \right)\left(\frac 13 + 2ky - 3x^2+3(4k^2-1)y^2 \right) \ge 0
 \end{equation}
 If $k=\frac 12$ the boundary is a parabola intersecting a line on the outsphere. 
 Otherwise the condition describes a cut hyperbola or a (possibly cut) ellipse, depending on the sign of $4k^2-1$:
 \begin{equation}
  \left(\frac 13 - 2ky\right)\left(3(4k^2-1)(y+\frac k{3(4k^2-1)})^2 - \frac{3k^2-1}{3(4k^2-1)} -3x^2 \right)\ge 0
 \end{equation}
If $k=0$ this is a circle, and if $k \to \frac 1{\sqrt{3}}$ the cut hyperbola goes into a 
triangle. A section from the second family is always related to an uncut ellipse.
 
 \item $a=0 \land (b=0 \lor c=0)$ (if we want to stay in the proper halfsimplex, the only possibility is $c=0$). These points form an edge of halfsimplex. The cubic condition reduces then to:
 \begin{equation} \label{fact2}
  \left( \frac 13 \pm x + k y \right) \left( \frac 13 + 3(k^2-1) y^2 \mp x \pm 6k xy  - ky \right) \ge 0
 \end{equation}
 If $k=0$ the condition describes a parabola intersected by a line:
 \begin{equation} \label{fact2p}
  \left( \frac 13 \pm x + k y \right) \left( \frac 13 - 3 y^2 \mp x \right) \ge 0
 \end{equation}
 If $k\ne 0$ the condition describes a hyperbola intersected by a line:
 \end{enumerate}
 \begin{equation} \label{fact2h}
  \left( \frac 13 \pm x + k y \right) \left( \left( kx \mp \frac 1{6k} \right)^2 - \left( (k^2-1)y \pm kx -k/6 \right)^2 + \frac {(1-3k^2)(1-k^2)}{36k^2}\right) \ge 0
 \end{equation}
If $k=\frac 1{\sqrt{3}}$ the cut hyperbola degenerates to a triangle.

\begin{fact} \label{sp-sec}
 Assume that the boundary of the section contains a line segment. Then the rest of boundary is a conic which connects to the segment in two exposed points.
\end{fact}

We now turn to the second question: if the boundary does not factorize, can it fail to be 
smooth? We will see that this happens if and only if the section contains a pure state. 

The cubic curve (\ref{cub-cond}) will fail to be smooth if and only if 

\begin{equation} \det{\rho} = \partial_x \det{\rho} = \partial_y 
\det{\rho} = 0 \ . \end{equation}

\noindent A straightforward calculation verifies that this happens if and only 
if 

\begin{equation} x^2+y^2 = \frac 13 \ . \end{equation}

\noindent This is precisely where the boundary of the 2-section touches the 
outsphere, of radius $\frac 1{\sqrt 3}$, which means that it happens if and 
only if the 2-section passes through a pure state. So we want to find all 
2-sections containing a pure state.

Assume that the matrix
\begin{equation} \label{sect-el}
\rho = I_3/3 + xA + yB =  \left[ \begin{array}{ccc} \frac 13 + x + ky & y a e^{i\phi} & y b e^{i\phi} \\
y a e^{-i\phi} & \frac 13 - x + ky & y c e^{i\phi} \\
y b e^{-i\phi} & y c e^{-i\phi} & \frac 13 - 2ky
\end{array} \right]
\end{equation}
is of rank one. We will use $X,Y,Z$ to denote the diagonal entries. First observe, that $y\ne 0$. Now consider the minor formed by the first and the second column and the first and the third row. We have that $Xcy e^{-i\phi}=aby^2$. If $\phi \not \in \{0,\pi\}$, then one has $a=0 \ \lor \ b=0$. In this case one can remove the phase, and it is enough to consider real matrices with $e^{i\phi}=\pm 1$.

We have three equations arising from non-main minors:
\begin{align*}
 & Xc=\pm aby & & Yb=\pm cay & & Za=\pm bcy
\end{align*}
If one of the numbers $a,b,c$ is zero, then at least one other has to be zero. In this case the matrix (\ref{sect-el}) has a block structure, the determinant factorises, and the pure state lies in the points where both factors simultaneously vanish. We have already discussed this 
case.

Observe that in the first family of sections with factorised boundary both factors can vanish simultaneously only if $k \le 1/2\sqrt{3}$. Otherwise the boundary is an ellipse without a pure state. In the limit case $k = 1/2\sqrt{3}$, one has an ellipse with one pure state on it.

\begin{figure}
 \centering
 \includegraphics[width=10cm]{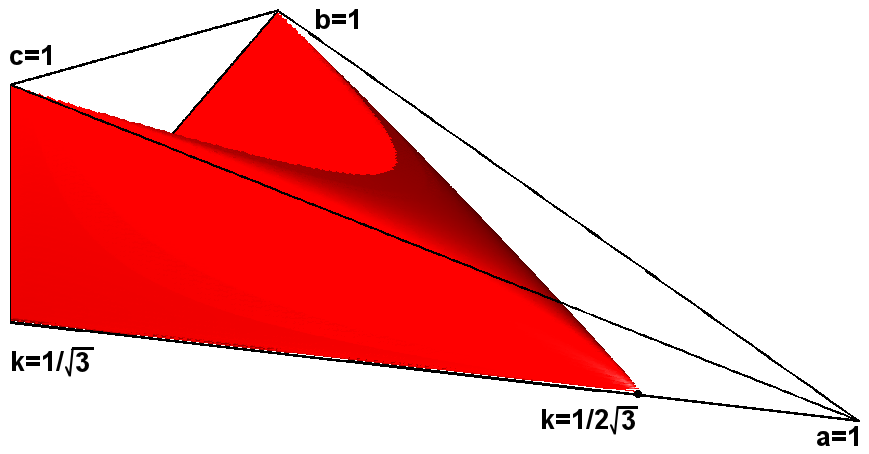}
 \caption{Points in the simplex $\varphi=0$ containing a pure state}
 \label{fig:unsmth}
\end{figure}

Consider now the case when $abc\ne 0$. Consider the equations for the main minors:
\begin{align*}
 & XY=a^2y^2 & & YZ=c^2y^2 & & ZX=b^2y^2
\end{align*}
One can easily calculate:
\begin{align*}
 & X=\pm \frac{bc}a y &&
 & Y=\pm \frac{ab}c y &&
 & Z=\pm \frac{ca}b y
\end{align*}
Using the normalisation of the trace one has $y = \frac{abc}{(ab)^2+(bc)^2+(ca)^2}$. 
Applying it to the above equations one gets:
$6kabc=(ab)^2+(ac)^2-2(bc)^2$. Points safisfying this equation lie on the surface presented on the figure \ref{fig:unsmth}.

An example of such a section is $k=0, a=b=c=\frac 1{\sqrt 3}$. The point representing this section lies in the middle of the upper wall of the simplex.

\section{Sections, projections, and numerical ranges} \label{duality}

An important property of the set of quantum states is its self-duality. This gives 
quantum logic its special flavour \cite{Mielnik, Wilce, Udu}. In the vector space 
of traceless hermitian $n\times n$ matrices the dual (or polar) $X^*$ of a set $X$ 
is defined by 

\begin{equation} X^* = \{ M: 1/n + \mbox{Tr}(MM') \geq 0 \ \forall \ M'\in X\} \ . 
\end{equation}

\noindent If we recall that any density matrix can be written as $\rho = I_n/n + M$, 
where $M$ is chosen so that $\rho$ is positive, it is easily seen that the self-duality 
of the set of density matrices follows from the fact that $\mbox{Tr}\rho \rho^\prime 
\geq 0$ for any pair of positive matrices $\rho$ and $\rho^\prime$. Subsets of the 
set of density matrices are not self-dual. If we take for example the set of separable states, the dual set will be a set of normalized entanglement witnesses. The properties of the duality operation are neatly summarized by 
\begin{align}
 & X^{**} = X \\
 & X \subset Y \iff X^* \supset Y^* \\
 & \emptyset^* = \mathbb{R}^{n^2-1} \\
 & (X \cup Y)^* = X^* \cap Y^* \\
 & (X \cap Y)^* = \mathrm{conv}(X^* \cup Y^*) \label{dual-int1}
\end{align}

\noindent where the convex hull appears in the last line. By definition the convex hull 
of a set is the smallest convex set that includes the given set. For a nonsingular linear transformation $A$ one has
 \begin{displaymath}
  (AX)^* = (A^{-1})^T X^*.
 \end{displaymath}
In particular if $A$ is orthogonal the set and its dual transform in the same way.

The reason why we bring this up here is that a cross-section of a self-dual body 
is dual to a projection onto the linear subspace defining the 
cross-section \cite{Weis}. Here we are interested in cross-sections using planes 
that pass through 
the origin (the maximally mixed state), and the projection is orthogonal (orthographic). 
Taking a cross-section by means of a plane not passing through the origin will 
give the dual of a perspective projection from a finite point. 
In mathematics a cross-section of a cone of 
semi-positive definite matrices is called a spectrahedron, and the question what 
kind of convex bodies that can be obtained as projections of spectrahedra arises 
naturally. Once we understand the set of 2-sections of the qutrit we can answer 
this question in our special case \cite{BWZ}.  

There is a further connection to the notion of numerical range of a matrix 
\cite{Gustafson, Dunkl2}. For a quadratic complex matrix $A$ we define a subset 
of complex plane called numerical range of $A$ by 
\begin{equation}
 \mathcal{W}(A)=\{ \mathrm{Tr}(A\rho): \rho \in \Omega \}
\end{equation}
where $\Omega$ denotes the set of density matrices. Let $A=aI+B+iC$, where $B,C$ are hermitian and traceless. Assume first that $B,C$ are orthogonal and of unit norm. Then one has
\begin{displaymath}
 \mathcal{W}(A)=\{ \rho_B + i \rho_C: \rho \in \Omega \} + a
\end{displaymath}
where $\rho_B$ and $\rho_C$ denote the components of $\rho$ respectively in the direction 
of $B$ and of $C$. We get a translated projection of $\Omega$ onto an affine subspace passing through the maximally mixed state. If we abandon the assumption that $B,C$ are normalized and orthogonal our set will be a linearly deformed and translated projection of $\Omega$. The numerical range of $3\times 3$ matrices is well understood \cite{Keeler}, 
and the idea was used recently to explore the shadows cast on 2-planes by the set of 
quantum states \cite{Dunkl1}. 

\section{Conclusions}

Our intention with this note was to give a parametrization of the set of {\sl all} 
2-sections of the qutrit. Theorem 1 is a quite satisfactory answer to this problem. 
In section \ref{GM} this enabled us to see at a glance how earlier works---in particular the interesting work by Goyal et al \cite{Goyal}---sample this set. In section 
\ref{gen-sec} we explored how the shape of the 2-section changes as we move through 
the set of all 2-sections, and in section \ref{duality} we gave a glimpse of a more general context to which the study of cross-sections of sets of positive matrices belongs.

\

\

\noindent \underline{Acknowledgements}: IB was supported by the Swedish Research Council 
under contract VR 621-2010-4060.

The study was supported by research fellowship within project ``Enhancing Educational Potential of Nicolaus Copernicus University in the Disciplines of Mathematical and Natural Sciences'' (project no. POKL.04.01.01-00-081/10.)

\end{document}